\documentclass{ws-ijmpe}

\begin{document}

\markboth{W. Trautmann {\it et al.}}{N/Z Dependence of Projectile Fragmentation 
}

\catchline{}{}{}{}{}

\title{N/Z DEPENDENCE OF PROJECTILE FRAGMENTATION 
}

\author{\footnotesize W. TRAUTMANN,$^{1}$
\footnote{E-mail: w.trautmann@gsi.de}
~P.~ADRICH,$^{1}$ 
T.~AUMANN,$^{1}$ 
C.O.~BACRI,$^{2}$ 
T.~BARCZYK,$^{3}$ 
R.~BASSINI,$^{4}$ 
S.~BIANCHIN,$^{1}$ 
C.~BOIANO,$^{4}$
A.S.~BOTVINA,$^{1,11}$ 
A.~BOUDARD,$^{5}$
J.~BRZYCHCZYK,$^{3}$  
A.~CHBIHI,$^{6}$ 
J.~CIBOR,$^{7}$ 
B.~CZECH,$^{7}$ 
M.~DE~NAPOLI,$^{8}$ 
J.-\'{E}.~DUCRET,$^{5}$ 
H.~EMLING,$^{1}$
J.D.~FRANKLAND,$^{6}$  
M.~HELLSTR\"{O}M,$^{1}$ 
D.~HENZLOVA,$^{1}$
G.~IMM\`{E},$^{8}$ 
I.~IORI,$^{4}$
H.~JOHANSSON,$^{1}$  
K.~KEZZAR,$^{1}$
A.~LAFRIAKH,$^{2}$ 
A.~LE~F\`EVRE,$^{1}$ 
E.~LE~GENTIL,$^{5}$ 
Y.~LEIFELS,$^{1}$
J.~L\"{U}HNING,$^{1}$ 
J.~{\L}UKASIK,$^{1,7}$ 
W.G.~LYNCH,$^{9}$
U.~LYNEN,$^{1}$ 
Z.~MAJKA,$^{3}$ 
M.~MOCKO,$^{9}$ 
W.F.J.~M\"{U}LLER,$^{1}$ 
A.~MYKULYAK,$^{10}$
H.~ORTH,$^{1}$
A.N.~OTTE $^{1}$  
R.~PALIT,$^{1}$ 
P.~PAW{\L}OWSKI,$^{7}$ 
A.~PULLIA,$^{4}$
G.~RACITI,$^{8}$, 
E.~RAPISARDA,$^{8}$ 
H.~SANN,$^{1,\ddagger}$ 
C.~SCHWARZ,$^{1}$ 
C.~SFIENTI,$^{1}$ 
H.~SIMON,$^{1}$
K.~S\"{U}MMERER,$^{1}$ 
M.B.~TSANG,$^{9}$
G.~VERDE,$^{9}$
C.~VOLANT,$^{5}$ 
M.~WALLACE,$^{9}$ 
H.~WEICK,$^{1}$
J.~WIECHULA,$^{1}$ 
A.~WIELOCH,$^{3}$  \and
B.~ZWIEGLI\'{N}SKI$^{10}$}

\address{$^{1}$ Gesellschaft f{\"u}r Schwerionenforschung mbH, Planckstr. 1, 
D-64291 Darmstadt, Germany\\
$^{2}$ Institut de Physique Nucl\'eaire, IN2P3-CNRS et Universit\'e, F-91406 Orsay, France\\
$^{3}$ M. Smoluchowski Institute of Physics, Jagiellonian University, Pl-30059 Krak\'ow, Poland\\
$^{4}$ Istituto di Scienze Fisiche, Universit\`a degli Studi and INFN, I-20133 Milano, Italy\\
$^{5}$ DAPNIA/SPhN, CEA/Saclay, F-91191 Gif-sur-Yvette, France\\
$^{6}$ GANIL, CEA et IN2P3-CNRS, F-14076 Caen, France\\
$^{7}$ IFJ-PAN, Pl-31342 Krak\'ow, Poland\\
$^{8}$ Dipartimento di Fisica e Astronomia dell'Universit\`a and INFN-LNS
and Sez. CT, I-95123 Catania, Italy\\
$^{9}$ Department of Physics and Astronomy and NSCL, MSU, East Lansing, MI 48824, USA\\
$^{10}$ A. So{\l}tan Institute for Nuclear Studies, Pl-00681 Warsaw, Poland\\
$^{11}$ Institute for Nuclear Research, Russian Academy of Science, Ru-117312 Moscow, Russia\\
$^{\ddagger}$ deceased}

\author{THE ALADIN 2000 COLLABORATION}


\maketitle

\begin{history}
\received{(received date)}
\revised{(revised date)}
\end{history}

\begin{abstract}
The $N/Z$ dependence of projectile fragmentation at relativistic energies has 
been studied in a recent experiment at the GSI laboratory with the ALADiN 
forward spectrometer coupled to the LAND neutron detector. 
Besides a primary beam of $^{124}$Sn, also secondary beams of $^{124}$La and $^{107}$Sn 
delivered by the FRS fragment separator have been used in order to extend the 
range of isotopic compositions of the produced spectator sources.
With the achieved mass resolution of $\Delta A/A \approx 1.5\%$, lighter isotopes with
atomic numbers $Z \le 10$ are individually resolved.
The presently ongoing analyses of the measured isotope yields focus on 
isoscaling and its relation to the properties of hot fragments at freeze-out and on  
the derivation of chemical freeze-out temperatures which are found to be independent of the 
isotopic composition of the studied systems. The latter result is at variance 
with the predictions for limiting temperatures as obtained with finite-temperature
Hartree-Fock calculations. 
\end{abstract}

\section{Introduction}

The present interest in isotopic effects in nuclear multifragmentation is
partly motivated by the importance of the nuclear equation of state for 
astrophysical processes.
Supernova simulations 
or neutron star models require inputs for the nuclear equation of state 
at extreme values of density and asymmetry for which the predictions 
differ widely.$^{1-5}$
Fragmentation reactions are of interest here because they permit 
the production of nuclear systems with subnuclear densities and 
temperatures which, e.g., largely overlap with those expected for the 
explosion stages of core-collapse supernovae.\cite{botv05} 
Laboratory studies of the properties of nuclear matter in the hot 
environment, similar to the astrophysical situation, are thus becoming feasible, 
and an active search for suitable observables 
is presently underway. Of particular interest is the 
density-dependent strength of the symmetry term which is essential for the 
description of neutron-rich objects up to the extremes encountered in neutron stars 
(refs.$^{3,7-9}$
and contributions to this workshop).

Here, new results from a recent experiment performed with the ALADIN spectrometer
at the GSI laboratory will be discussed in which the possibility 
of using secondary beams for reaction studies at relativistic energies 
has been explored. Beams of $^{107}$Sn, $^{124}$Sn, $^{124}$La, and $^{197}$Au 
as well as Sn and Au targets
were used to investigate the mass and isospin dependence of projectile 
fragmentation at 600 MeV per nucleon. The neutron-poor radioactive 
projectiles $^{107}$Sn and $^{124}$La were produced at the Fragment Separator FRS
by fragmentation of a primary beam of $^{142}$Nd and delivered to the ALADIN 
experiment.

The results presented in the following, after a brief discussion of global observables, 
will be restricted to quantities deduced from the measured yields of resolved isotopes.
They will, in particular, include isoscaling and its relation to the properties of hot 
fragments at the low-density freeze-out, as well as chemical temperatures deduced
from double ratios of isotopic yields and the consequences for the limiting-temperature 
concept in multifragmentation. Preliminary results from the present 
experiments have been presented previously.$^{10-12}$

\begin{figure}[htb]     

   \centerline{\includegraphics[height=5.5cm]{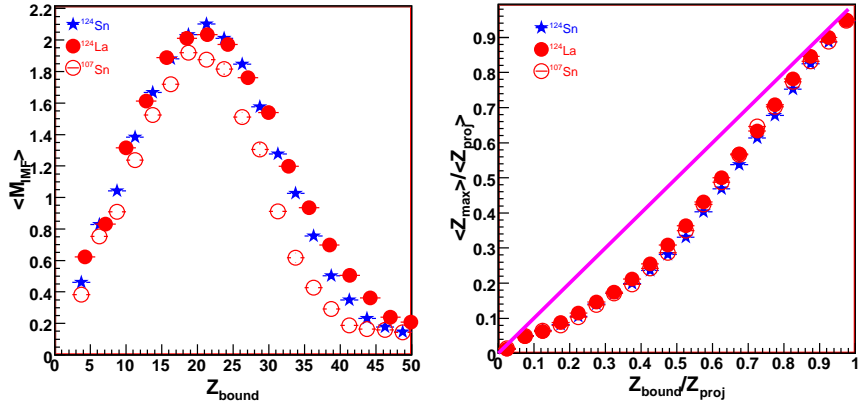}}
\caption{Mean multiplicity $<$$M_{\rm IMF}$$>$ of intermediate-mass fragments 
$3 \le Z \le 20$ produced in the fragmentation of $^{107,124}$Sn and $^{124}$La
at 600 A MeV as a function of $Z_{\rm bound}$ (left panel). The right 
panel shows the correlations of the mean $Z$ of the largest fragment with 
$Z_{\rm bound}$ with both quantities being normalized with respect to the 
projectile $Z$.
}
\label{global} 
\end{figure}

\section{N/Z dependence of global fragmentation observables}

The dependence on the isotopic composition of the system is rather small for global 
observables of the studied reactions. This is shown in Fig.~\ref{global} for the 
multiplicity of the produced intermediate-mass fragments and for $Z_{\rm max}$,
both as a function of $Z_{\rm bound}$. Here
$Z_{\rm max}$ denotes the largest atomic number $Z$ within a partition while 
the sorting variable $Z_{\rm bound} = \Sigma Z_i$ with $Z_i \ge 2$ 
is related to the impact parameter 
and inversely correlated with the degree of excitation of the produced 
spectator system. The multiplicities exhibit the universal
rise and fall of fragment production\cite{schuett96} (Fig.~\ref{global}, left panel), 
and only a slightly steeper slope in the rise
section distinguishes the neutron-rich case of $^{124}$Sn from the other two 
systems. As confirmed by statistical model calculations, this difference is related 
to the evaporation properties of excited heavy nuclei.\cite{sfienti_prag} 
Neutron emission prevails for neutron-rich nuclei which leads to a concentration 
of the residue channels, with small associated fragment multiplicities, 
in a somewhat narrower range of large $Z_{\rm bound}$ than in the case of the 
neutron-poor systems. The evaporation of protons reduces $Z_{\rm bound}$ since protons 
are not counted therein.

This effect is, nevertheless, small and nearly invisible
in the correlation of $<$$Z_{\rm max}$$>$ with $Z_{\rm bound}$ (Fig.~\ref{global}, 
right panel). There, the transition from predominantly residue production 
to multifragmentation becomes apparent as a reduction of $<$$Z_{\rm max}$$>$ with
respect to $Z_{\rm bound}$
which occurs between $Z_{\rm bound}/Z_{\rm proj} =$~0.6 and 0.8. 
The range of isotopic compositions of the studied nuclei $N/Z = 1.14$~to 1.48
is not large enough for significant variations of the reaction 
mechanism to appear. A shrinking of the coexistence zone in the 
temperature-density plane is predicted for neutron-rich matter but observable 
consequences can only be expected for asymmetries
far beyond those presently available for laboratory studies.\cite{muell95,mekjian01}
This fact, on the other hand, has the important consequence that the basic reaction 
process is the same for all the present systems, a prerequisite for the interpretation 
of isoscaling and its relation with the symmetry energy. This point will be further 
quantified in the section on temperatures.

\section{N/Z dependence of light fragment production}

The mass resolution obtained for projectile fragments entering into the 
acceptance of the ALADIN spectrometer is about 3\% for fragments with $Z \le 3$
and decreases to 1.5\% for $Z\geq 6$.
Masses are thus individually resolved for fragments with atomic number $Z \leq 10$.
The elements are resolved over the full range of atomic numbers up 
to the projectile $Z$ with a resolution of $\Delta Z \leq 0.2$ obtained with the
TP-MUSIC IV detector.\cite{sfienti_prag}

\begin{figure}[htb]	
   \centerline{\includegraphics[width=0.85\textwidth]{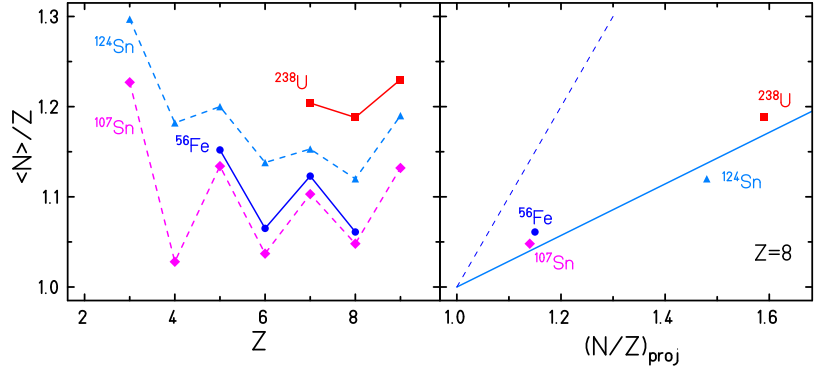}}
\caption{Inclusive mean values $<$$N$$>$/$Z$ of light fragments with $3 \le Z \le 9$ 
produced in the fragmentation of $^{107,124}$Sn (600 A MeV, ALADIN),
$^{56}$Fe and $^{238}$U (both 1 A GeV, FRS, from refs.\protect\cite{ricci04,napo04}) 
as a function of the fragment $Z$ (left panel). 
The right panel shows the results for $Z=8$ 
as a function of the $N/Z$ value of the projectile. The lines represent the 
trend of the data (full line) and $<$$N$$>$/$Z = (N/Z)_{\rm proj}$ (dashed).
}
\label{noverz} 
\end{figure}

The inclusive mean neutron-to-proton ratio $<$$N$$>$/$Z$ of light 
fragments in the range $3 \le Z \le 9$ is presented in Fig.~\ref{noverz}. 
The values for $Z=4$ have been corrected for the missing yield of unstable $^8$Be 
by including an estimate for it obtained from a smooth interpolation over the 
identified yields of $^{7,9-11}$Be. 
This correction has a negligible effect for the case 
of $^{107}$Sn because $<$$N$$>$/$Z$ of the detected beryllium isotopes 
is close to 1 already. For $^{124}$Sn the value
of $<$$N$$>$/$Z$ for $Z=4$ is lowered from 1.24 to 1.18 which makes the systematic odd-even 
variation more clearly visible in the neutron rich case. 
The odd-even staggering is, however, much more pronounced in the case of 
the neutron-poor $^{107}$Sn.
The strongly bound even-even nuclei attract a large fraction of the product 
yields during the secondary evaporation stage.\cite{ricci04} This effect is,
apparently, larger if the hot fragments are already close to symmetry, as it is expected 
for the fragmentation of $^{107}$Sn.\cite{buyuk05}

Inclusive data\cite{ricci04,napo04} obtained with the FRS fragment separator at GSI for 
$^{238}$U and $^{56}$Fe fragmentations on titanium targets
at 1 A GeV bombarding energy confirm that the observed patterns 
are very systematic~(Fig.~\ref{noverz}, left panel).
Nuclear structure effects characteristic for the isotopes produced combine with
significant memory effects of the isotopic composition of the excited 
system by which they are emitted. This has the consequence that, because of its 
strong variation with $Z$, averaged neutron-to-proton ratios $<$$N$$>$/$Z$ for
unspecified ranges of $Z$ are not very useful for studying nuclear matter properties. 
For this purpose, techniques will have to be used which cause the nuclear structure 
effects to cancel out. Selecting a particular element, e.g. $Z=8$, is already sufficient
to reveal the linear correlation of $<$$N$$>$/$Z$ with the $N/Z$ of the projectile 
(Fig.~\ref{noverz}, right panel). According to the Statistical Fragmentation 
Model (SMM\cite{bond95}), the correlation should be much stronger for the
excited fragments at the breakup stage and, in fact, not deviate much from the
dashed $<$$N$$>$/$Z = (N/Z)_{\rm proj}$ line in the figure.\cite{buyuk05} 
The reduction is due to sequential decay and its tendency of directing the isotope
distributions closer to the valley of stability. For a quantitative analysis, 
a precise modeling of these secondary processes will, therefore, be required. 

\begin{figure}[htb]           
\centering
\begin{minipage}[c]{.45\textwidth}
\vspace{-0.3mm}
   \centerline{\includegraphics[height=7.45cm]{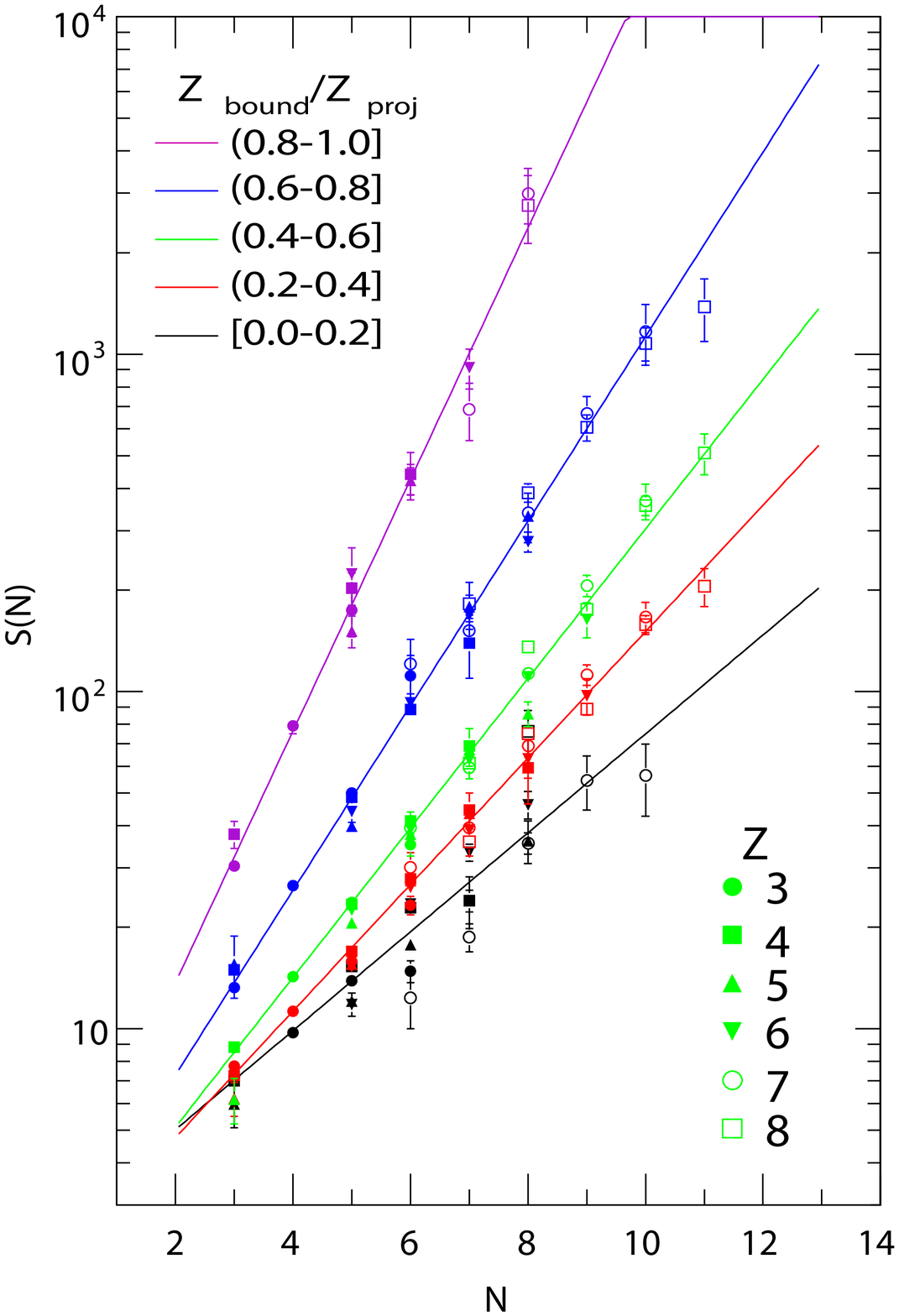}}
\vspace{0.3mm}
\end{minipage}
\begin{minipage}[c]{.53\textwidth}
   \centerline{\includegraphics[height=7.4cm]{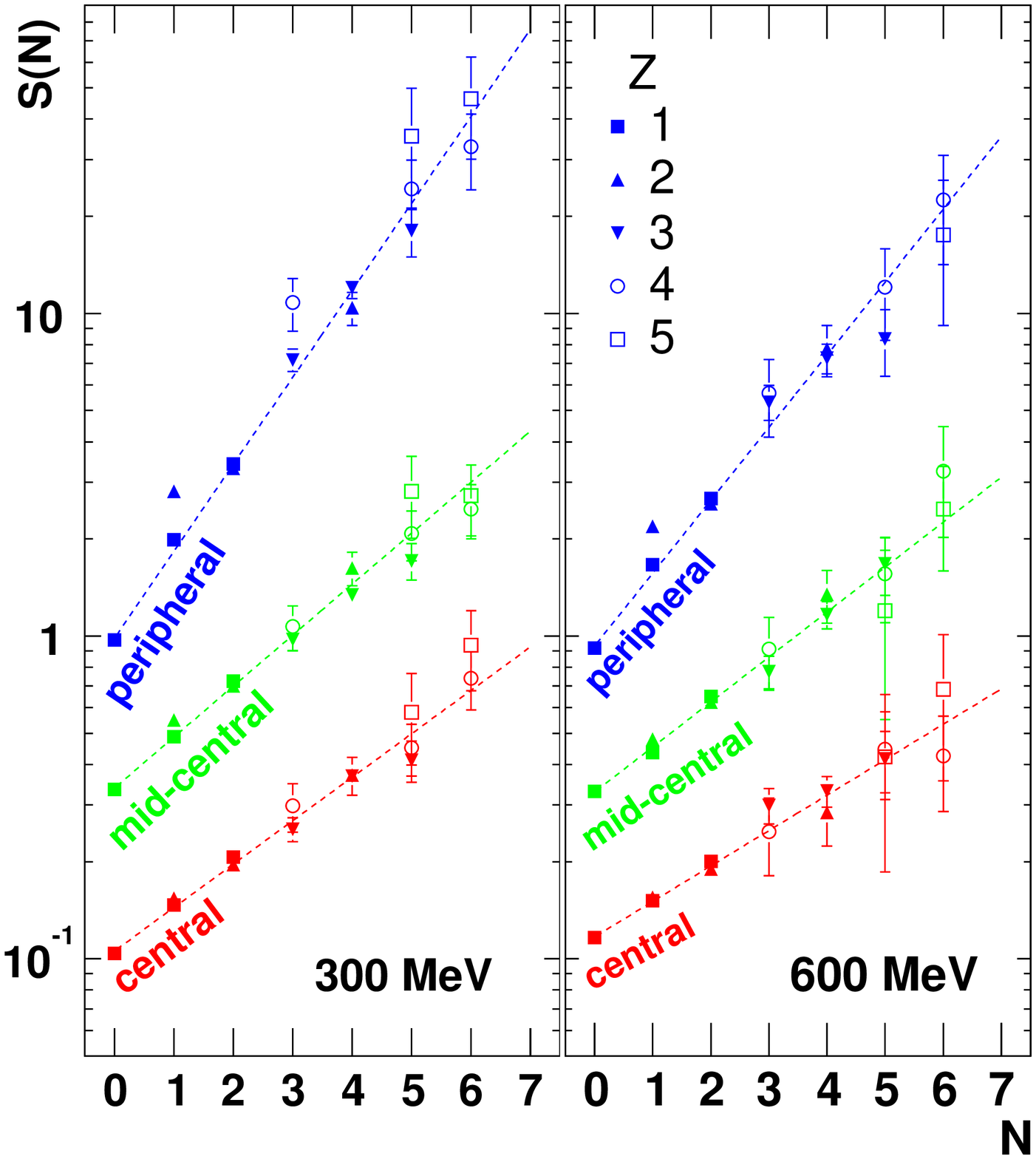}}
\end{minipage}
\caption{Scaled isotopic ratios $S(N)$ for Li to O isotopes from the
fragmentation of $^{124}$Sn and $^{107}$Sn projectiles at 600 A MeV (ALADIN, left panel)
and for H to B isotopes from $^{12}$C + $^{112,124}$Sn at $E/A$ = 300 and 600~MeV 
(INDRA, right panel, from ref.\protect\cite{lef05}) as a function of the neutron 
number $N$. The centrality selection 
was obtained with $Z_{\rm bound}$ (ALADIN, left panel, five bins as indicated) and
according to charged particle multiplicity as measured with the full detector
(INDRA, right panel, three bins with ''central'' indicating $b/b_{\rm max} \leq 0.4$).
The dashed lines in both panels are the results of exponential fits according 
to Eq.~(\protect\ref{eq:scalab}). Only statistical errors are displayed; note also
the offset factors of multiples of three in the right panel.
}
\label{isoscal} 
\end{figure}

\section{Isoscaling}

\begin{figure}[htb]    

   \centerline{\includegraphics[height=5.5cm]{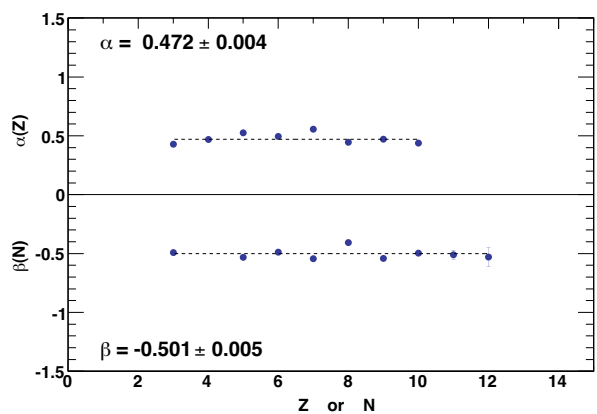}}
\caption{Isoscaling parameters $\alpha (Z)$ and $\beta (N)$ obtained by individually
fitting the $N$ and $Z$ dependences of isotopes and isotones, respectively,
for inclusive yields from the fragmentation 
of $^{124}$Sn and $^{107}$Sn projectiles at 600 A MeV. 
The dashed lines represent the mean values for isotopes with $Z$ from 3 to 10 
and isotones with $N$ from 3 to 12. Only statistical errors are displayed.
}
\label{individual} 
\end{figure}

The phenomenon of isoscaling has been shown 
to be common to many different types of heavy-ion 
reactions.$^{20-23}$
It is observed by comparing product yields
$Y_i$ from reactions which differ only in the isotopic
composition of the projectiles or targets or both. 
Isoscaling refers to an
exponential dependence of the measured yield ratios $R_{21}(N,Z)$
on the neutron number $N$ and proton number $Z$ of the detected 
products. The scaling expression
\begin{equation}
R_{21}(N,Z) = Y_2(N,Z)/Y_1(N,Z) = C \cdot exp(\alpha N + \beta Z)
\label{eq:scalab}
\end{equation}
describes rather well the measured ratios over a wide range of
complex particles and light fragments.\cite{tsang01a} 
The accuracy with which the isoscaling relation is obeyed is conveniently judged 
by regarding the scaled isotopic ratios $S(N) = R_{21}(N,Z)/{\rm exp}(\beta Z)$, 
obtained by dividing out the common $Z$ dependence after fitting 
according to eq.~(\ref{eq:scalab}). Isoscaling means that all the scaled yield ratios 
will follow the same exponential dependence on the neutron number $N$.
The results for the pair of reactions with $^{124}$Sn and $^{107}$Sn 
projectiles, sorted into five bins of $Z_{\rm bound}$ are shown in 
Fig.~\ref{isoscal}, left panel. 
Alternatively, one may also regard the behavior of the exponential slopes 
$\alpha (Z)$ of the yield ratios for individual elements or $\beta (N)$ for 
individual chains of isotones. As demonstrated in Fig.~\ref{individual} for the inclusive
yields from the same pair of reactions, 
the deviations from the common mean value are impressively small.
Besides the precision with which isoscaling is observed in the range of fragments from 
Li to Ne, also the monotonic decrease of the slope of $S(N)$ with centrality represents 
a remarkable observation. Qualitatively, it resembles that observed for the reactions
$^{12}$C + $^{112,124}$Sn at 300 and 600~MeV per nucleon, 
studied with INDRA at GSI (ref.\cite{lef05}, right panels of Fig.~\ref{isoscal}).

The statistical model offers a simple physical 
explanation for the appearance of isoscaling in finite systems.
Charge distributions of fragments with fixed mass numbers $A$, as well 
as mass distributions for fixed $Z$, are approximately Gaussian with 
average values and variances which are connected with the temperature,
the symmetry-term coefficient, and other parameters. 
(Here the symmetry-energy term, $E_{\rm sym} = \gamma (A-2Z)^2/A$ with the
coefficient $\gamma$, is that of the liquid-drop description of the nascent
fragments at freeze-out and $\gamma = 25$~MeV is normally used.\cite{bond95})
The mean values of the fragment distributions depend 
on the total mass and charge of the systems, e.g. via the chemical 
potentials in the grand-canonical approximation, while 
the variances depend mainly on the physical conditions reached,
the temperature, the density and possibly other variables. For example, 
the charge variance $\sigma_Z\approx \sqrt(AT/8\gamma)$ obtained for
fragments with a given mass number $A$ 
is only a function of the temperature and of the symmetry-term coefficient 
since the Coulomb contribution is very small.\cite{botv85}  This relation of isoscaling
with the symmetry energy has attracted considerable interest 
recently.\cite{botv02,soul03,lef05,ono03,shetty07}

In the grand-canonical approximation,
assuming that the temperature $T$ is about the same (see below),
the scaling parameters $\alpha$ and $\beta$ are given by 
the differences of the neutron and proton chemical potentials for
the two systems divided by the temperature, 
$\alpha = \Delta \mu_{\rm n}/T$ and $\beta = \Delta \mu_{\rm p}/T$.
The proportionality of $\Delta \mu_{\rm n}$ and thus of 
the isoscaling parameters with the coefficient $\gamma$ of
the symmetry-energy term 
has been obtained from the statistical 
interpretation of isoscaling within the SMM\cite{botv02} and 
Expanding-Emitting-Source Model\cite{tsang01a} 
and confirmed by an analysis of reaction dynamics.\cite{ono03} 
The relation is
\begin{equation} \label{eq:dmunu}
\Delta \mu_{\rm n} = \mu_{\rm n,2} - \mu_{\rm n,1} \approx 4\gamma
(\frac{Z_{1}^2}{A_{1}^2}-\frac{Z_{2}^2}{A_{2}^2}) = 4\gamma \Delta (Z^2/A^2)
\end{equation}
where $Z_{1}$,$A_{1}$ and $Z_{2}$,$A_{2}$ are the charges and mass
numbers of the neutron-poor and neutron-rich systems, respectively, at breakup. 
The difference of the chemical potentials
depends essentially only on the coefficient $\gamma$ of the symmetry term
and on the isotopic compositions. 
A proportionality of $\alpha$ and $1/T$, with $T$ derived from double isotope ratios, 
has been observed for light-ion (p, d, $\alpha$) induced reactions at bombarding energies 
in the GeV range.\cite{botv02} Also the symmetry-term coefficient $\gamma \approx 22.5$~MeV,
deduced according to eq.~(\ref{eq:dmunu}), supported the statistical approach.
These data were inclusive and a value close to the standard value was to be expected
because the fragment multiplicities are small in this case.\cite{beaulieu} 

The present reactions studied with ALADIN cover the full rise and fall 
of fragment production while the $^{12}$C induced reactions studied with INDRA
cover mainly the rise up to $Z_{\rm bound} \approx Z_{\rm proj}/2$ with multiple
fragment production being dominant in central collisions.\cite{schuett96}
In both cases, the measured temperatures rise slightly with increasing centrality 
but not sufficiently fast in order to compensate for the decrease of the 
isoscaling parameter $\alpha$. The product $\Delta \mu_{\rm n} = \alpha \cdot T$, 
therefore, decreases.
With the assumption that the isotopic compositions remain close to those of the 
original $^{112,124}$Sn targets, the evolution of the apparent symmetry-term coefficient 
$\gamma_{\rm app}$, i.e. without corrections for the effects of sequential decay, 
was followed into the regime of multifragmentation with the $^{12}$C induced data 
measured with INDRA.\cite{lef05} It was again found to be close to 
the normal-density coefficient $\gamma \approx 25$~MeV for peripheral collisions,
associated with small fragment multiplicities, but dropped to lower values of 
15 to 17 MeV for the central impact parameters leading to multifragmentation.
This result is confirmed by the preliminary analysis of the new data 
for projectile fragmentation measured with ALADIN. The product $\alpha \cdot T$ drops
by about 20\% through the rise of fragment production but then 
remains fairly constant once the regime
of multifragmentation has been reached ($Z_{\rm bound} /Z_{\rm proj} \le 0.5$).

\begin{figure}[htb]     
   \centerline{\includegraphics[height=6.0cm]{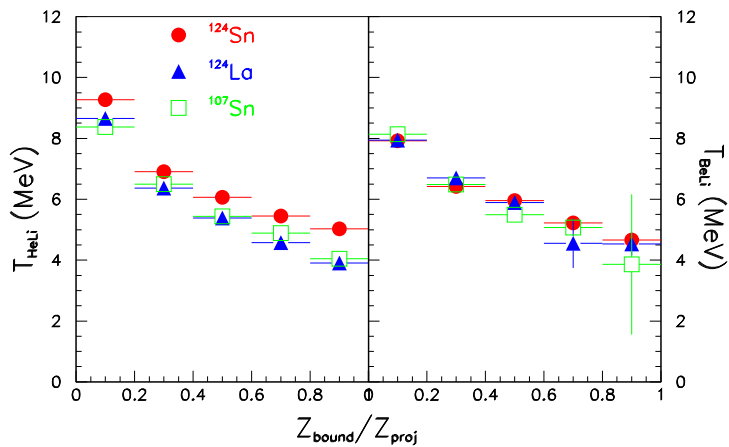}}

\caption{Apparent temperatures $T_{\rm HeLi}$ (left panel) and $T_{\rm BeLi}$ 
(right panel) as a function of $Z_{\rm bound}$ for the three reaction systems
produced with $^{107,124}$Sn and $^{124}$La projectiles. 
Only statistical errors are displayed.
}
\label{temp} 
\end{figure}

\section{N/Z dependence of the nuclear caloric curve}

Two examples of double-isotope temperatures deduced from the measured isotope yields 
are shown in Fig.~\ref{temp} as a function of $Z_{\rm bound}$. Besides the frequently
used $T_{\rm HeLi}$ (left panel), determined from $^{3,4}$He and $^{6,7}$Li yields also 
the results for $T_{\rm BeLi}$ are displayed (right panel). For $T_{\rm BeLi}$ the 
isotope pairs of $^{7,9}$Be and $^{6,8}$Li are used which each differ by two neutrons. The 
double difference of their binding energies amounts to 11.3 MeV and is nearly as large 
as the 13.3 MeV in the $T_{\rm HeLi}$ case. The apparent temperatures are displayed,
i.e. no corrections for secondary decays feeding the ground states of these nuclei 
are applied.

Both temperature observables show the same smooth rise with increasing centrality that
was observed earlier in a study of $^{197}$Au fragmentations.\cite{traut07} 
According to $T_{\rm BeLi}$, the temperatures at the chemical freeze-out are
identical for all three reaction systems
and have the same dependence on $Z_{\rm bound}$.
This is not equally visible in $T_{\rm HeLi}$ which exhibits slightly larger values
for the neutron rich case of $^{124}$Sn. Inspection of the single isotope ratios shows, 
however, that the $^{3,4}$He ratio represents an exception in that it varies much less
than the other ratios with the neutron-richness of the system. A possible
explanation is offered by the observations shown in Fig.~\ref{noverz}. The strong
population of even-$N$-even-$Z$ nuclei, evident from the behavior of $<$$N$$>$/$Z$, 
suggests that also $^{4}$He nuclei are abundantly produced. The  
effect is stronger in the neutron-poor cases for which the apparent $T_{\rm HeLi}$ 
will be accordingly lower. Taking this nuclear-structure effect into account 
is likely to reduce the differences between systems also for this thermometer.

The observed invariance of the breakup temperature with the isotopic composition of 
the system is of interest because it is opposite to what is expected according to the 
finite-temperature Hartree-Fock calculations of Besprosvany and Levit.\cite{besp89}
The limiting temperatures obtained for the stability of excited 
nuclei are strongly dependent on the Coulomb pressure generated by the protons they contain.
Along isobars, the limiting temperatures decrease rapidly toward the proton rich side and, 
along the valley 
of stability, they decrease with increasing mass because the effect of the increasing 
$Z$ dominates over that of the decreasing $Z/A$. For the nuclei studied here, a difference 
of about 2 MeV is predicted between the limiting temperature of $^{124}$Sn 
and those for the 
proton rich $^{107}$Sn and $^{124}$La or for the heavier $^{197}$Au.\cite{besp89}
A difference of this magnitude is clearly not seen in the data (Fig.~\ref{temp}).
Neither does a comparison with the apparent temperatures measured for $^{197}$Au 
fragmentations 
(Fig.~7 in ref.\cite{traut07}) provide evidence for the predicted mass dependence.
The temperatures are the same, within errors, when regarded as a function of the scaled 
quantity $Z_{\rm bound} / Z_{\rm proj}$. This, however, is suggested by the 
$Z_{\rm bound}$ scaling of fragmentation observables if a comparison of sources of 
different mass but equal excitation is intended to be made.\cite{schuett96}

\section{Discussion}

It is one of the basic assumptions of isotopic reaction studies that the mechanism as such
remains unchanged and that effects related to the asymmetry dependence of the nuclear forces
can be isolated by changing nothing but the isotopic composition of the system. This view is
supported by the observed invariance of the chemical breakup temperatures, here for the 
case of spectator fragmentation. The interpretation of the breakup temperatures
as a manifestation of the limiting temperatures predicted by the Hartree-Fock 
model,\cite{besp89} on the other hand, is not equally supported.
This interpretation was derived from the $Z_{\rm bound}$ dependence of the temperature 
which suggested a mass dependence because of the changing mass of the spectator 
system.\cite{nato95}
Later on, the predicted mass dependence was found to be confirmed in a compilation of 
chemical break-up temperatures from several experiments,\cite{nato02} 
including the earlier ALADIN 
data, which led to their use as a general reference for testing
predictions of nuclear stability with respect to temperature 
(see ref.\cite{kelic06} for references). The new data for the $A \approx 120$ region and 
their comparison with $^{197}$Au fragmentations suggest that the observed variation
with decreasing $Z_{\rm bound}$ is more related to the increasing excitation energy 
than to the decreasing mass of the produced spectator system.
Breakup temperatures fairly independent of the isotopic composition or mass of the system 
are, e.g.,
predicted by the statistical fragmentation models which are based on the concept 
of a phase-space driven instability.\cite{ogul02} 

The discussion of the isoscaling results has focussed on mainly three topics, among them 
primarily the effects of sequential decay for the isoscaling parameters and for the 
symmetry term, as well as the role of the surface term in the symmetry energy.
Surface terms gain in importance as the mean mass of the produced fragments 
decreases in multi-fragment decays. Thirdly, also a possible modification of the 
source $N/Z$ due to preequilibrium emissions prior to breakup will affect and
most probably reduce the isoscaling parameters. So far, however, none of these mechanisms
has been identified as providing an obvious explanation for the large drop that is observed. 
Calculations with the microcanonical Markov-chain version of the SMM,\cite{botv01}
even show that sequential decay will cause the isoscaling coefficients to increase
if the symmetry term
governing the mass distributions of the hot fragments is reduced.\cite{lef05}
Transport models predict that the isotopic composition at breakup
should not deviate by more than a few percent from the original value.\cite{gait04}
New experimental results, furthermore, indicate that the isotopic dependence of the 
surface term is reduced, with respect to that needed to describe ground-state masses, as the 
excitation of the system increases.\cite{surface06} This is also expected 
theoretically.\cite{ono03}

A decrease of the isoscaling parameter $\alpha$ with increasing violence
of the collision, beyond that expected from the simultaneous increase of the
temperature, has also been observed for reactions at intermediate energy
(refs.\cite{shetty07,souliotis06} and references therein) and a systematics 
is emerging. A very recent interpretation of these results 
arrives at the conclusion that the reduced density, rather than the elevated 
temperature, causes the symmetry term to be smaller than the standard 
value.\cite{bao06} This analysis uses a temperature independent potential part 
of the symmetry term and a kinetic part in the form of a Fermi gas, and it is argued
that the isoscaling phenomenon itself is evidence for the fact that properties of 
the considered homogeneous nuclear matter are observable in fragmentation reactions.

In the Statistical Multifragmentation Model,\cite{bond95} normal-density fragments 
statistically distributed within an expanded volume are considered. 
Here a reduction of the symmetry term in their liquid-drop description 
can be imagined to be caused by fragment 
modifications in the hot environment, including deformations or the effects of
nuclear interactions between them. 
The neglect of such effects in the actual codes should probably be regarded 
as an idealization as the assumption of equilibrium at the final freeze-out stage
requires nuclear interactions in order to achieve it. 
The consequences of a reduced symmetry term for predicted global fragment observables,
as e.g. multiplicity, are rather small for this class of models. The 
partitioning is predominantly driven by the surface term in the fragment description
while a variation of the symmetry term affects mainly the isotopic 
distributions.\cite{buyuk05,ogul02,surface06}

Further studies will be needed in order to put some of the assumptions on a firmer basis 
and to quantitatively establish the reduction of the symmetry term in multi-fragment 
breakups. The sequential decay corrections are obviously
important but also the evolution of the isotopic compositions during the reaction
process deserves attention. It will have to be explored whether an 
experimental reconstruction of the
neutron-to-proton ratio of the detected spectator systems will be feasible using
the coincident neutron data measured with the LAND detector.\cite{iwm05}

This work has been supported by the European Community under contract No. HPRI-CT-1999-00001, 
by the Polish State Committee for Scientific Research (KBN) under 
contract No. 2P03B11023, and by the
Polish Ministry of Science and Higher Education under Contracts No. 1 P03B 105
28 (2005 - 2006) and N202 160 32/4308 (2007-2009).

\end{document}